\documentclass[epj]{svjour}

\usepackage{bm}
\usepackage{hyperref}
\usepackage{amssymb}
\usepackage{amsopn}
\usepackage{subfigure}
\usepackage{graphicx}

\newcommand{\be}{\begin{equation}}
  \newcommand{\ee}{\end{equation}}
\newcommand{\ba}{\begin{eqnarray}}
  \newcommand{\ea}{\end{eqnarray}}
\newcommand{\Mp}{m_p}

\newcommand{\ka}{\kappa}
\newcommand{\de}{\delta}
\newcommand{\De}{\Delta}

\newcommand{\Lam}{\Lambda}
\newcommand{\bx}{{\bf x}}
\newcommand{\bk}{{\bf k}}
\newcommand{\phio}{\phi_0}
\newcommand{\cs}{c_s}
\newcommand{\al}{\alpha}
\newcommand{\vp}{\varphi}

\newcommand{\xo}{X_0}

\newcommand{\calc}{{\cal A}}
\newcommand{\calp}{{\cal P}}
\newcommand{\vsg}{\bar{\sigma}}
\newcommand{\cg}{\tilde{\cal C}}

\renewcommand{\(}{\left(}
\renewcommand{\)}{\right)}
\renewcommand{\[}{\left[} 
\renewcommand{\]}{\right]}

\begin{document}

\title{CMB constraints on noncommutative geometry during inflation}
\author{Khamphee Karwan
\thanks{Email:pk\_karwan@yahoo.com}
}
\institute{
Department of Physics, Faculty of Science, Kasetsart University, Bangkok 10900, Thailand
}

\abstract{
We investigate the primordial power spectrum of the density perturbations
based on the assumption that  space is noncommutative in the early stage of inflation, and
 constrain the contribution from noncommutative geometry using CMB data.
Due to the noncommutative geometry, the primordial power spectrum can lose rotational invariance.
Using the k-inflation model and slow-roll approximation, we show that
the deviation from rotational invariance of the primordial power spectrum
depends on the size of noncommutative length scale $L_s$ but not on sound speed.
We constrain the contributions from the noncommutative geometry to the covariance matrix 
of the harmonic coefficients of the CMB anisotropies
using five-year WMAP CMB maps.
We find that the  upper bound for $L_s$ depends on the product of sound speed and slow-roll parameter.
Estimating this product using cosmological parameters from the five-year WMAP results,
the upper bound for $L_s$ is estimated to be less than $10^{-27}{\rm cm}$ at 99.7$\%$ confidence level.
} 


\maketitle

\section{Introduction}

The inflationary cosmology \cite{Guth:81, Linde:82, Albrecht:82} is the scenario of the very early universe.
It provides a successful mechanism for generating nearly scale invariant primordial density perturbations,
that give rise to galaxy formation and temperature anisotropies in the CMB which are in agreement with observation \cite{wmap5:cosmos}.
If the period of inflation is sufficiently longer than that required for solving the horizon and flatness problems,
such that the wavelengths of perturbations which are observed today emerged from the Planck regime in the early stages of inflation,
the physics on trans-Planckian scales should leave an imprint on the primordial density perturbations \cite{Brand:01, Martin:01}.
Here, we consider the imprint of trans-Planckian physics based on noncommutative spacetime.

Near the Planck scale, the properties of spacetime
are expected to be modified due to the quantum nature of gravity \cite{Szabo:03}.
It has been shown that a consequence of string theory
which is a promising candidate of quantum gravity, is that the 
spacetime is noncommutative \cite{Seiberg:99}
\be
\[ x^\mu ,x^\nu \] = i\Theta^{\mu\nu}(x),
\label{ncm}
\ee
where $\Theta^{\mu \nu}$ is an antisymmetric tensor.

The influences of spacetime noncommutativity on the feature of 
power spectrum of primordial fluctuations have been studied by many authors \cite{Brand:02} - \cite{Joseph2:08}.
For the case where $\Theta^{i j} = 0$ but $\Theta^{0 i} \neq 0$ \cite{Brand:02, Tsujikawa:03, Huang:03},
the contribution from spacetime noncommutativity can lead to the running
of the spectral index of the primordial power spectrum.
For the case where $\Theta^{i j} \neq 0$ but $\Theta^{0 i} = 0$ \cite{Lizzi:02, Fang:08},
the primordial power spectrum can become direction-dependent, and consequently
the statistics of CMB fluctuations becomes anisotropic.
We are interested in this noncommutative geometry induced statistical anisotropy.

Usually, the statistics of the CMB temperature fluctuations is supposed to be isotropic.
Hence, if the non-Gaussianity of the CMB fluctuations is assumed to be negligible,
the statistical properties of the CMB fluctuations will be completely described by the
angular power spectrum \cite{Picon:06}.
However, recently there are many attempts to check whether the statistics of the
CMB fluctuations is perfectly isotropic by searching for the statistical anisotropy
contributions in the CMB sky maps \cite{Hajian:06, Picon:08, Groeneboom:09}.
In the case where the statistics of the CMB fluctuations is anisotropic,
the angular power spectrum does not contain all the information about the statistical properties of the CMB fluctuations
even when the Gaussianity of the CMB fluctuations is assumed.
Some of the estimators for quantifying the statistical anisotropy
contributions in the CMB fluctuations have been proposed in \cite{Hajian:03, Picon:06, Pullen:07}.
According to \cite{Hajian:06, Picon:08}, the statistics of the observed CMB fluctuations
does not deviate from isotropy significantly.

In this work, we constrain the contributions from noncommutative geometry 
to CMB temperature fluctuations using five-year WMAP CMB maps.
In the next section, we compute the power spectrum of primordial density perturbations
by taking the spacetime noncommutativity of the form  $\Theta^{i j} \neq 0$ and $\Theta^{0 i} = 0$.
In section 3, we compute the covariance matrix for the harmonic coefficients of the CMB temperature fluctuations
$\langle a^*_{lm} a_{l'm'}\rangle$,
and constrain the contributions from noncommutative geometry using CMB maps.
Finally, we conclude in section 4.

\section{The contributions from noncommutative geometry}

In this section, we investigate the contributions from noncommutative geometry to
the primordial power spectrum in the k-inflation model \cite{Picon:99, Garriga:99}.

\subsection{The second order action of perturbation}

We start with the general action of the inflaton of the form
\be
S = \frac 12 \int d^{4}x\sqrt{-g}\[R + 2P(X,\phi) \],
\label{act1}
\ee
where $\phi$ is the inflaton field and
$X = -(1/2) \partial_\mu\phi\partial^\mu\phi$.
Here, we have set the reduced Planck mass $(8\pi G)^{-1/2} = 1$.
To study the evolution of density perturbations during inflation,
one expands the action (\ref{act1}) around the homogeneous and isotropic background.
In our consideration, we use the ADM metric formalism in which the
line element is given by \cite{Fang:08, Maldacena:02, Chen:07}
\be
d^2s=-N^2d^2t+h_{ij}(dx^i+N^idt)(dx^j+N^jdt).
\ee
Using this line element, the action (\ref{act1}) takes the form
\ba
S &=& \frac 12 \int d^4x\sqrt{h} N(R^{(3)}+2P)
\label{act2}
\\
& &+\frac 12 \int d^4x\sqrt{h}
N^{-1}(E_{ij}E^{ij}-E^{2}),
\nonumber
\ea
where $h=\det(h_{ij})$,
$E_{ij}=\frac{1}{2}(\dot{h}_{ij}-\nabla_{i}N_{j}-\nabla_{j}N_{i})$,
$E=E_{i}^{i}$, a dot denotes the derivative with respect to time
and $\nabla_{i}$ is the covariant derivative compatible with $h_{ij}$.
The three-dimensional Ricci curvature $R^{(3)}$ is
computed from the metric $h_{ij}$.

In the ADM formulation $h_{ij}$ and $\phi$ are the dynamical variables, while $N$ and $N^i$
are Lagrange multipliers. To compute the perturbed action in
the slow-roll approximation, it is convenient to use the uniform curvature gauge
in which \cite{Maldacena:02}
\be
\delta \phi \equiv \vp(t,\bx) \quad {\rm and } \quad 
h_{ij} = a^2 \delta_{ij}, 
\ee
where $a$ is the cosmic scale factor, $\delta \phi (t,\bx)= \phi (t,\bx) - \phio(t)$ is the perturbation in the inflaton field
and the subscript $0$ represents the background value.

In order to find the second order action for $\vp$,
we  first compute the constraint equations for $N$ and $N^i$
from the action (\ref{act2}) and solve these equations for $N$ and $N^i$
to first order of $\vp$. the result is \cite{Fang:08}
\be
N=1+\alpha \quad {\rm and} \quad N_i = \partial_i \psi.
\label{n}
\ee
Up to the lowest order of slow-roll parameter, the parameters $\al$ and $\psi$ can be written as
\be
\alpha = \frac{H\epsilon}{\dot\phi_0}\vp \quad {\rm and} \quad
\partial^{2}\psi =(\frac{P_{,\phi_0}}{2H}-\frac{3H^2}{\dot\phi_0}\epsilon)\vp - \frac{H\epsilon}{\dot\phi_0\cs^2}\dot\vp,
\ee
where $H = \dot a/a$ is the Hubble parameter,
$\epsilon = - \dot H / H^2 = X P_{,X}/H^2$ is the slow-roll parameter,
$\cs^2 = P_{,X}/\rho_{,X}$ is the sound speed,
the subscript $,X$ denotes a derivative with respect to $X$
and $\rho$ is the energy density of the inflaton.

Substituting equation (\ref{n}) back in the action and
expanding the action to the second order of perturbation, we obtain
\ba
\de S^{(2)} &=& \int d^4x a^3(1 + \al ) 
\biggl[ P_{,\phio}\vp + P_{,\phio\phio}\frac{\vp^2}{2}
+P_{,\xo} \biggl( \dot\vp\dot\phio
\nonumber\\
&&
- (2\al - 3\al^2)\xo - N^i\partial_i\vp\dot\phio - \frac{\partial_\mu\vp\partial^\mu\vp}{2} - 2\al\dot\vp\dot\phio \biggr) 
\nonumber\\
&&
+ \frac 12 P_{,\xo\xo}
\biggl(4\al^2\xo^2 - 4\al\xo\dot\vp\dot\phio + (\dot\vp\dot\phio)^2\biggr) + \dots\biggr]
\nonumber\\
&&
+ \frac 12 \int d^4x a^3(1+\al)^{-1}(-6H^2) + \dots .
\label{act-1}
\ea
Keeping  the lowest order of slow-roll parameter and second order of $\vp$, the above action becomes \cite{Fang:08}
\ba
\de S^{(2)} &=& \int
d^4x\frac{a^3}{2}\(P_{,\xo\xo}(\dot\phio\dot\vp)^2 - P_{,\xo}\partial_\mu\vp\partial^\mu\vp\),
\label{act-21}\\
&=& \int
d^4x\frac{a^3}{2}\(\frac{P_{,\xo}}{\cs^2}\dot\varphi^2-P_{,\xo}(\partial\varphi)^2\).
\label{act-22}
\ea
Here, we have used $\rho_{,X} = P_{,X} + 2P_{,XX}X$.
From the above calculation, we see that in the slow-roll approximation the second order action of the
field perturbation in uniform curvature gauge can be obtained by just expanding
Lagrangian $P(X,\phi)$ around the homogeneous field $\phio$.
This is because the terms that are multiplied by the metric perturbation
in the perturbed action are subleading in slow-roll parameter \cite{Fang:08, Maldacena:02}.

\subsection{noncommutative geometry}

We now study how the noncommutative geometry influences the action of the field perturbations.
In order to take the effect of noncommutative geometry into account, we replace the ordinary products
in the action with the star products.
In curved spacetime, the star product can be expanded as \cite{Lizzi:02}
\ba
f \star g &\equiv & \sum_{k=0}^{\infty} \frac{1}{k!} \left( \frac{i}{2} \right)^k\Theta^{\mu_1 \nu_1} \cdots \Theta^{\mu_k \nu_k} 
\times
\nonumber\\
&&
 \left( D_{\mu_1}\cdots D_{\mu_k} f \right)
\left( D_{\nu_1} \cdots D_{\nu_k} g\right), 
\label{star}
\ea
where $D_{\mu}$ is the covariant derivative.
In this work, we consider the case where $\Theta^{0i} = 0$ while $\Theta^{ij} \neq 0$.
The case where $\Theta^{ij} \neq 0$ can arise in the effective theory of D-branes.
In some model of D-branes, e.g., 
a Dp-brane on p-dimensional Torus with unequal radii $R_i$ \cite{Chu:99}, the non-commutative parameter $\Theta$ is not necessary equal in all direction.
This implies that a preferred direction can occur in space-space noncommutativity.
We would like to investigate some effects of this form of isotropy breaking by
following the literature \cite{Lizzi:02,Fang:08}
to choose, Without losing generality, a special frame in which the
nonzero components of $\Theta^{\mu\nu}$
are $\Theta^{12} = - \Theta^{21} = L_s^2/a^2$, where
$L_s$ is the noncommutative length scale \cite{Lizzi:02}.

According to the previous section, the terms in the second order action that are multiplied by metric perturbation
are subleading in slow-roll parameter. Hence, the noncommutative effect can be 
incorporated by replacing the products between fields in the Lagrangian by the star products, and expanding
the Lagrangian to the second order in the field perturbation.
In the calculation of star product, we also ignore the metric perturbation because it gives rise to the terms that are subleading in slow-roll parameter.
Since it is possible to study the effect of noncommutative geometry perturbatively \cite{Lizzi:02},
we do the calculation up to the lowest non trivial order in $\Theta^{\mu \nu}$.
Up to the second order in $\Theta^{\mu\nu}$, the star products
of the multiple functions can be written as \cite{Fang:08}
\ba
f_1\star \cdots \star f_n &=& \biggl(1
+ \frac{i}{2}\Theta^{\mu\nu}\sum_{a<b}D_\mu^aD_\nu^b
\label{star2}\\
&&
-\frac{1}{8}\Theta^{\mu\nu}\Theta^{\rho\sigma}
\sum_{a<b,c<d}D_\mu^aD_\nu^bD_\rho^cD_\sigma^d \biggr)
f_1\cdots f_n,
\nonumber
\ea
where $a,b,c,d$ run over $1,\dots ,n$.

For illustration, we suppose that $P(X,\phi)$ can be written as $P(X,\phi) = F(\phi)G(X) = \phi^n X^m$,
and use eq. (\ref{star2}) to compute the star product between field varibles.
We first consider the case where the derivatives in eq. (\ref{star2}) act on the terms in $G(X)$ only, such that
$\de_\Theta P(X,\phi) = F(\phi)\de_\Theta G(x)$, where $\de_\Theta$ denotes the contribution from noncommutative geometry.
In this case, the contribution from noncommutative geometry starts to appear at the second order of $\Theta^{\mu\nu}$, i.e.,
\ba
\de_\Theta G &=& 
-\frac{1}{8}\Theta^{\mu\nu}\Theta^{\rho\sigma}
\sum_{a<b,c<d}D_\mu^aD_\nu^bD_\rho^cD_\sigma^d G
\nonumber\\
&=& -\frac{1}{8}\Theta^{\mu\nu}\Theta^{\rho\sigma}\sum_{a<b,c<d}D_\mu^aD_\nu^bD_\rho^cD_\sigma^d(-\frac 12 \partial_\gamma\phi\partial^\gamma\phi)^m,
\nonumber\\
&=& -\frac{1}{8}\Theta^{\mu\nu}\Theta^{\rho\sigma} \biggl(
\frac{m(m-1)}{2}\xo^{m-2} D_\mu D_\rho X_1 D_\nu D_\sigma X_1
\nonumber\\
&&
- \frac{m}{2}\xo^{m-1}D_\mu D_\rho \partial_\gamma\vp D_\nu D_\sigma \partial^\gamma\vp
\nonumber\\
&&
- m(m-1)\xo^{m-2} X_1 D_\mu D_\rho \partial_\gamma\vp D_\nu D_\sigma \partial^\gamma\phio
\biggr),
\nonumber\\
&=& 
-\frac{1}{8F(\phio)}\Theta^{\mu\nu}\Theta^{\rho\sigma} \biggl(
\frac 12 P_{,\xo\xo} D_\mu D_\rho X_1 D_\nu D_\sigma X_1
\nonumber\\
&&
- \frac 12 P_{,\xo} D_\mu D_\rho \partial_\gamma\vp D_\nu D_\sigma \partial^\gamma\vp
\nonumber\\
&&
- P_{,\xo\xo} X_1 D_\mu D_\rho \partial_\gamma\vp D_\nu D_\sigma \partial^\gamma\phio\biggr),
\label{deg}
\ea
where $X_1 = -\partial_0\phio\partial^0\vp$ is the firse order perturbation in $X$.
In this equation, we omit the terms that are proportional to $D_\mu D_\rho\partial_\gamma\phio D_\nu D_\sigma\partial^\gamma\phio$,
such as $P_{,\xo\xo}X D_\mu D_\rho\partial_\gamma\phio D_\nu D_\sigma\partial^\gamma\phio$, because these terms vanish after integration by parts.
Using  similar consideration, one can show that if the derivatives in eq. (\ref{star2}) act on the terms in $F(\phi)$
only, the noncommutative contribution will be proportional to $P_{,\phio\phio}$,
and if the derivatives act on the terms in both $F(\phi)$ and $G(X)$,
the noncommutative contribution will be proportional to $P_{,\phio\xo}$.
In our case, both contributions can be neglected because the terms that are proporttional to $P_{,\phio\phio}$ and $P_{,\phio \xo}$
are subleading in slow-roll parameter.
Hence, the noncommutative geometry modifies the action in leading order of slow-roll parameter as
\be
\de_\Theta S^{(2)} = \de S^{(2)} + \de_\Theta S_X^{(2)},
\label{act-non}
\ee
where $\de S^{(2)}$ is the action for the perturbed field in eq. (\ref{act-22})
and 
\be
\de_\Theta S_X^{(2)} = \int d^4x a^3 F(\phio)\de_\Theta G.
\label{act-non1}
\ee
The action $\de_\Theta S_X^{(2)}$ can be spit into three parts as
$\de_\Theta S_X^{(2)} = \de_\Theta S_{X1} + \de_\Theta S_{X2} + \de_\Theta S_{X3}$,
where the subscripts $1, 2,3$ denote the contribution from the first, second and third terms on the RHS of eq. (\ref{deg}).
The expression for $\de_\Theta G$ in eq. (\ref{deg}) is valid for a generic $P (X,\phi)$,
because one always do Taylor's expansion of $P(X,\phi)$ around the background 
when applying the star product in the action for perturbation.
Moreover, the action (\ref{act-non}) can be obtained by just replacing products with  star products in the action (\ref{act-21}).

We suppose that $D_\rho\Theta^{\mu\nu} = 0$, and
do integration by parts for the action (\ref{act-non1}) in the similar way as \cite{Lizzi:02}.
After doing integration by parts and evalulating the covariant derivatives, each part of $\de_\Theta S_X^{(2)}$ becomes
\ba
\de_\Theta S_{X1} 
&=& \int d^4 x\frac{aH^2L_s^4}{8} P_{,XX} X \partial_p\partial_0\vp\partial_p\partial^0\vp,
\\
\de_\Theta S_{X2} &=& \int d^4 x\frac{aH^2L_s^4}{16} P_X\biggl(
\partial_p \partial_\mu \vp\partial_p \partial^\mu\vp
\nonumber\\
&&
+ 2H\partial_p \dot\vp \partial_p\vp
- H^2 \partial_p \vp \partial_p\vp
\biggr),
\\
\de_\Theta S_{X3} &=& -\int d^4 x\frac{aH^2L_s^4}{8} P_{,XX} 2X\partial_p\partial_0\vp (\partial_p\partial^0\vp + H\partial_p\vp),
\nonumber
\ea
where $p = 1,2$.
In the above three equations and following calculation, we omit the subscript $0$  for $P$, $\rho$ and $X$
because from now on we will treat them as the background quantities.
We note that in the calculation of $\de_\Theta S_{X3}$ we suppose that $\dot X < X$.
due to the slow-roll approximation.
From these results, we get
\ba
\de_\Theta S_X^{(2)} &=& \frac 1{16} \int d^4x aH^2L_s^4\rho_{,X}
\biggl[ \beta \partial_p \partial_0 \vp\partial_p \partial^0\vp
\label{actstar}\\
&&
+ \cs^2\partial_p \partial_i \vp \partial_p\partial^i\vp
+ 2\beta H \partial_p \dot\vp \partial_p\vp
- \cs^2 H^2 \partial_p \vp \partial_p\vp 
\biggr],
\nonumber
\ea
where $\beta = 2\cs^2 - 1$.
In order to find the evolution equation and the corresponding solution from action (\ref{act-non}),
it is convenient to write the action in terms of the variable $v = z\zeta$,
where $z = a\sqrt{2X\rho_{,X}}/H$ \cite{Garriga:99}.
The variable $\zeta$ is the curvature perturbation in comoving gauge which has a direct connection with
the generation of large scale structure and CMB fluctuations.
This quantity is related to $\vp$ through  $\zeta = H\vp / \dot\phio$.
Expressing the action (\ref{act-non}) in terms of $v$
and expanding $v$ in Fourier space as
\be
v(t, \bx) = 
\int \frac{d^3\bk}{\(2\pi\)^{3/2}}
\( {\bf a_k} v_{\bf k}(t) {\rm e}^{i\bk{\cdot} \bx} + h.c. \),
\ee
where ${\bf a_k},{\bf a_k^\dagger}$ satisfy the canonical commutation
relations, the action (\ref{act-non}), evaluated in the vacuum after normal ordering, becomes
\ba
S &=& \int d\eta d^3\bk\Biggl[ \vert v_{\bf k}' \vert^2
+ \( \frac{a''}{a} - \cs^2k^2 \) \vert v_{\bf k} \vert^2
\nonumber \\
&& 
-\frac{L_s^4H^2 k_\perp^2}{8a^2}
\biggl( \beta\vert v_{\bf k}' \vert^2 - \cs^2k^2 \vert v_{\bf k} \vert^2
\label{ncmact}\\
&&
+ \(\beta\(6\frac{a''}{a} - 10(Ha)^2\)
- \frac{\beta - 1}{2}(Ha)^2\)\vert v_{\bf k} \vert^2 \biggr) \Biggr],
\nonumber
\ea
where the prime denotes  derivative with respect to the conformal time $\eta = \int d a / a$ and
$k_\perp^2 = k_1^2 + k_2^2$.
In the above action, we have neglected the terms $\dot\cs^2$ and $\dot\rho_{,X}$
because these terms are subleading in slow-roll parameter.
Up to the lowest order of slow-roll parameter, we can write
$a^2 = 1/(H\eta)^2 + {\cal O}(\epsilon) \simeq 1/(H\eta)^2$ and $a'' /a \simeq 2/\eta$
Using these approximations and writing $v_{\bf k}$ in terms of the new variable
\be
y_{\bk}(\eta) = (1 - \beta\ka^2\eta^2 k_\perp^2)^{1/2} v_{\bk}(\eta),
\ee
where $\ka^2 = H^4L_s^4 / 8$ represents the contribution from noncommutative geometry,
the action (\ref{ncmact}) up to first order in $\ka^2$ takes the simple form
\ba
S &=& \int d \eta d^3\bk \biggl[ \vert y_{\bk}' \vert^2
- \(\cs^2 k^2 
-\frac{2}{\eta^2}\)\vert y_{\bf k} \vert^2
\nonumber\\
&&
+ \ka^2k_\perp^2\(\beta 
- \frac{\beta - 1}{2}\)
\vert y_{\bf k} \vert^2 \biggr].
\label{acty}
\ea
In the derivation of this action, we suppose that $\ka^2k_\perp^2\eta^2 < 1$ and $|\cs^2|$ do not  differ much from unity,
so that one can expand $1/(1-\beta\ka^2k_\perp^2\eta^2) \simeq 1 + \beta\ka^2k_\perp^2\eta^2 + \dots$
and neglect the terms that are proportional to $\beta\ka^2k_\perp^2\eta^2$.
This implies that this action is valid only if
\be
\eta^2 < \eta_i^2 \equiv \frac{8}{H^4L_s^4 k^2}.
\ee
For a suitable choice of initial conditions, we have $- \infty < \eta < 0$ during inflation,
so that the perturbation mode $k$ exits the horizon 
at $\eta_c = - 1/k$ and exits the sound horizon at $\eta_{cs} = - 1/(k\cs)$.
Hence, we will be able to follow the evolution of the perturbation mode $k$
well before its horizon (sound horizon) exit, i.e., $\eta_i < \eta_c$ ($\eta_i < \eta_{cs}$),
if the noncommutative length scale is required to be smaller than the Hubble radius, $L_s < H^{-1}$.

\subsection{Power spectrum}

In order to study the influence of noncommutative geometry on the behavior of density perturbation,
we compute the power spectrum of the curvature perturbation $\zeta$ from action (\ref{acty}).
From this action, one can show that the evolution equation for $y_{\bk}$ is
\be
y_{\bk}^{\prime \prime } + \( \cs^2k^2\gamma^2 - \frac{2}{\eta^2}\) y_{\bk} = 0,
\label{eqy}
\ee
where $\gamma^2 = 1 + \ka^2\sin^2(\theta)$
and $\theta = \sin^{-1}(k / k_\perp)$ denotes the angle between the vectors $\bk$ and $\hat{k}_3$.
Following  standard procedure, the solution of eq. (\ref{eqy}) is given by \cite{Riotto:02}
\be
y_{\bk} = \frac{e^{-ik\cs\gamma\eta}}{\sqrt{2k\cs\gamma}}
\( 1 + \frac{i}{k\cs\gamma\eta}\right),
\label{ysol}
\ee
so that the curvature perturbation is given by
\be
|\zeta_{\bk}| = \left | \frac{v_{\bk}}{z}\right| \simeq
\frac{\sqrt{2\pi} H}{\Mp\sqrt{k^3\cs\epsilon}}\(1 - \frac 34 \ka^2\sin^2(\theta)\),
\ee
and finally we obtain the primordial power spectrum
\be
{\cal P}_{\bk}^{\zeta }
= 
\frac{H^2}{\pi\Mp^2\cs\epsilon}\(1 - \frac 32 \ka^2\sin^2(\theta)\),
\label{prizeta}
\ee
where $\Mp = G^{-1/2}$ is the Planck mass.
It can be seen that the obtained power spectrum is direction-dependent due to the noncommutative space.
The magnitude of the noncommutative contribution depends on the angle between the vectors $\bk$ and $\hat{k}_3$,
but not on the sound speed.
From the expression for $\ka$, it can be seen that the deviation from rotational invariance of the power spectrum
also depends on the noncommutative length scale and Hubble parameter during inflation.
One can check that this deviation from rotational invariance
is similar to the one that is computed from canonical inflaton field \cite{Lizzi:02}.
This implies that, in the slow-roll approximation, the effect of noncommutative geometry on primordial power spectrum
is the same for both standard inflation and k-inflation model.
We note that the slow-roll approximation is required in our calculation because
when space is noncommutative, it is not easy to obtain the analytic expression for the primordial power spectrum
The amplitude of this power spectrum can be defined as $A_s = H^2/(\pi\Mp^2\cs\epsilon )$.
This quantity as well as $\ka$ are evaluated when the perturbation mode $k$ exits the horizon.
It can be seen that if we neglect the contributions from noncommutative geometry, the above power spectrum gives
rise to the usual power spectrum as in \cite{Garriga:99}.

\section{The CMB constraints}

In this section, we will constrain the contributions from noncommutative geometry
to CMB anisotropies using the CMB data.
It is well known that if the primordial power spectrum is direction-dependent,
the statistics of CMB will become anisotropic, i.e., the two-point function of the temperature fluctuations is
no longer rotationally invariant. Here, we  consider only the two-point function because we assume
the non-Gaussianity of the CMB fluctuations to be negligible.
In addition to noncommutative geometry, the direction-dependent primordial power spectrum can also be a consequence
of many phenomena in the early universe, for example see \cite{Ackerman:07, Mota1:08, Mota2:08}.
To compare theoretical prediction with the observation, it is convenient to expand the temperature anisotropy into spherical harmonics 
\be
\De T(\hat{n}) = T(\hat{n}) - T_0 = \sum_{l,m} a_{lm}Y_{lm}(\hat{n}),
\ee
where $T_0$ is the mean temperature of the CMB.
If we also expand the deviation from rotational invariance into  spherical harmonics,
the direction-dependent primordial power spectrum
can be written as \cite{Ackerman:07, Pullen:07}
\be
{\cal P}(\bk) = A(k) 
\[1 + \sum_{LM} \calp_{LM}Y_{LM}(\hat{k})\].
\label{prip}
\ee
Using this primordial power spectrum,
the covariance matrix of $a_{lm}$ can be written as
\be
\langle a_{l_1 m_1} a_{l_2 m_2}^* \rangle =
\de_{l_1 l_2}\de_{m_1 m_2} C_{l_1}
 + \sum_{LM} \Xi_{l_2 m_2 L M}^{l_1 m_1} D^{LM}_{l_1 l_2}.
\label{corr}
\ee
Here, $C_{l_1}$ is the CMB angular power spectrum, given by
\be
C_{l_1} = (4\pi)^2\int_0^\infty
 dk\,k^2A(k)[T_{l_1}(k)]^2,
\ee
where $T_{l_1}$ is the CMB transfer function which is taken to be rotationally invariant.
It is known that if the two-point function of the temperature anisotropy is not rotationally invariant,
the covariance matrix of $a_{lm}$ will not be diagonal.
The off diagonal elements of the covariance matrix appear in the second term of eq. (\ref{corr})
This term is given by
\be
D^{LM}_{l_1 l_2} = (4\pi)^2(-i)^{l_1 -l_2}\int_0^\infty
dk\, k^2 A(k) \calp_{LM} T_{l_1}(k)T_{l_2}(k),
\ee
and
\be
\Xi_{l_1 m_1 l_2 m_2}^{l_3 m_3} =
\sqrt{\frac{ (2l_1 +1)(2l_2 +1)}{ 4 \pi (2l_3+1) }}
\cg_{l_10l_20}^{l_30}
\cg_{l_1m_1l_2m_2}^{l_3m_3},
\ee
where $\cg_{l_1m_1l_2m_2}^{l_3m_3}$ are the Clebsch-Gordan coefficients.
Comparing the power spectrum in eq. (\ref{prip}) with the one in eq. (\ref{prizeta}),
we find that the non zero components of $\calp_{LM}$ are
\be
\calp_{0 0} = -\frac{4\sqrt{\pi}}{3} \frac 32 \ka^2 \quad {\rm and} \quad
\calp_{20} = \frac 43 \sqrt{\frac{\pi}{5}} \frac 32 \ka^2,
\label{g}
\ee
We see that the contributions from noncommutative geometry also influence the diagonal elements of the covariance matrix, i.e.
they modify the amplitude of $C_l$.
Substituting $\calp_{LM}$ from eq. (\ref{g}) into eq. (\ref{corr}),
we obtain the covariance matrix
\ba
\langle a_{l_1 m_1} a_{l_2 m_2}^* \rangle && \simeq 
C_{l_1} \Biggl\{
\de_{l_1 l_2}\de_{m_1 m_2} \biggl[ 1
+ \frac{3\ka^2}{2}
\times
\nonumber\\
&&
\biggl(
\frac{l_1^2-m_1^2}{4 l_1^2 - 1}
+ \frac{ ( l_1+1 )^2 -m_1^2}{ (2 l_1 + 3)(2 l_1 + 1)}
- 1
\biggr) \biggr]
\nonumber\\
&&
- \biggl( 
\de_{l_1 -2 l_2} \de_{m_1 m_2}
\frac{3\ka^2C_{l_1 l_2}}{2 (2l_1 -1 )}
\times
\label{cnoi}\\
&&
 \sqrt{ \frac{ ( (l_1 -1)^2 -m_1^2 ) ( l_1^2-m_1^2 )}{( 2 l_1 - 3 ) ( 2 l_1 + 1 )}}
  + l_1 \leftrightarrow l_2 \biggr)
\Biggr\},
\nonumber
\ea
where $C_{l_1 l_2} = (4\pi)^2\int_0^\infty dk k^2 A(k) T_{l_1}(k)T_{l_2}(k) / C_{l_1}$.

We next constrain the contributions from noncommutative geometry in this covariance matrix
using the five-year WMAP foreground-reduced maps \cite{wmap5:map}.
Since the noncommutative geometry  also influences the diagonal elements of the covariance matrix
and its contribution depends on $H^4$,
we write the noncommutative contribution in terms of the amplitude of the primordial power spectrum
$A_s$ as
\be
\frac 32 \ka^2  = A_s^2 \Sigma,
\ee
where
\be
\Sigma = \frac{3}{16} \pi^2 \Mp^4 \epsilon^2\cs^2 L_s^4.
\label{sigma}.
\ee
We adopt the procedures in \cite{Picon:08} to compute the posterior probability of the
parameters $\Sigma$ and $A_s$, given the observed temperature anisotropies $\mathbf{a}$,
\be
P(\vsg | \mathbf{a}) \propto L(\mathbf{a} | \vsg ) P(\vsg),
\ee
where $\vsg = \{\Sigma, A_s\}$ is the set of parameters,
$L(\mathbf{a} | \vsg )$ is the likelihood and $P(\vsg)$ is the prior.
Since the galactic contamination cannot be completely removed from some regions of the sky,
one does not have the full-sky CMB maps with well-defined error properties.
To reduce the galactic contamination, one masks the contaminated regions as
\be
c_i = M_i \De T_i,
\ee
where $i$ is the pixel index, $c_i$ is the masked CMB map, $\De T_i$ is the full-sky map
and $M_i$ is a mask which is zero at the contaminated points and is one elsewhere.
The above relation can be written in harmonic space as
\be
c_{lm} = M_{lm, l'm'}b_{l'm'},
\label{clm}
\ee
where the matrix $M_{lm, l'm'}$ is given by
\be
M_{l m,l' m'} = \sum_{L M} M_{L M} \Xi_{L M l' m'}^{l m}.
\ee
Here, $c_{lm}$, $M_{lm}$ and $b_{lm}$ are the spherical harmonic coefficients of $c_i$, $M_i$ and $\De T_i$ respectively.
Moreover, due to the instrument noise, the finite beam resolution and the discreteness of the temperature maps,
the contributions to the unmasked CMB map come from the sum of the instrument noise with the convolution
between the signal of the CMB anisotropies and the window function, such that
\be
\De\mathcal{T} = W \mathbf{a} + \mathcal{N},
\label{winn}
\ee
where $W$ is the window function and $\mathcal{N}$ is the instrument noise.
We suppose that the contributions from the beam  and the pixel asymmetries are negligible.
Using eqs. (\ref{clm}) and (\ref{winn}), the covariance matrix of the masked temperature multipoles can be written as
\ba
C_{lm, l'm'} = \sum_{l_1 m_1 l_2 m_2} && M_{lm, l_1m_1} \biggl[ W_{l_1}\langle a_{l_1 m_1} a_{l_2 m_2}^*\rangle W_{l_2}
\nonumber\\
&&
 + N_{l_1 m_1, l_2 m_2} \biggr] M_{l_2m_2, l'm'}^* ,
\label{covat}
\ea
where $N_{l_1 m_1, l_2 m_2}$ is the pixel noise covariance matrix, given by
\be
N_{l_1 m_1, l_2 m_2} = \De a \sum_{L M} \Xi_{L M l_2 m_2}^{l_1 m_1} N_{L M}.
\ee
Here, $\De a$ is the area of each pixel in the temperature map,
and $N_{L M}$ is defined as
\be
N_{L M} = \sum_{i} \De a \frac{\sigma_0^2}{n_i^{\rm obs}} Y_{L M}(\hat{r}_i),
\ee
where $\sigma_0$ is the rms noise of a single observation and
$n_i^{\rm obs}$ is the number of observations of pixel $i$.

In order to compute the likelihood, the inversion of the covariance matrix is required.
Since the inversion of the large matrix is time consuming,
we avoid to inverse the large covariance matrix by writing the likelihood
function in terms of the reduced bipolar coefficients instead of $c_{lm}$.
The reduced bipolar coefficients are defined as \cite{Hajian:06}
\be
A_{L M} = \sum_{l_1 m_1 l_2 m_2}\,\,
(-1)^{m_2} d_{l_1m_1}d_{l_2m_2}^* 
\cg_{l_1m_1l_2 \,-m_2}^{LM},
\ee
where $d_{lm}$ are the harmonic coefficients of the temperature anisotropies.
For the full-sky and noiseless case, the mean  of the reduced bipolar coefficients
for the noncommutative k-inflation can be computed using eq. (\ref{cnoi}), and
the non zero components are
\ba
\langle A_{00} \rangle &=& \sum_l (-1)^l C_l\sqrt{2l +1}\(1 - \frac{2}{3}A_s^2\Sigma\),
\\
\langle A_{20} \rangle &=& -2A_s^2\Sigma\sum_l (-1)^l C_l \sqrt{\frac{l (l + 1)(2l + 1)}{45 (2l -1)(2l + 3)}} 
\nonumber\\
&&
- 2A_s^2\Sigma\sum_{l\geq 4} (-1)^l C_l \sqrt{\frac{2 l (l - 1)}{15 (2l -1)}}.
\label{alm}
\ea
In the more realistic case, the mask and the instrument noise must be taken into account,
so that we compute the reduced bipolar coefficients using eqs. (\ref{covat})  and (\ref{cnoi}).
The result is
\be
\langle A_{LM} \rangle = \sum_{l_1 l_2 m_1 m_2} (-1)^{m_2} 
C_{l_1m_1, l_2m_2}
\cg_{l_1m_1l_2\, -m_2}^{LM}.
\label{meanb}
\ee
It can be seen from the above equation that due to the effect of the mask, $\langle A_{00}\rangle$
and $\langle A_{20}\rangle$ will not be the only  non zero
components of $\langle A_{LM}\rangle$. 

We define
\be
\de A_{LM} = A_{LM} - \langle A_{LM} \rangle ,
\label{delalm}
\ee
and write the covariance matrix for $\de A_{LM}$ as
\ba
&& \calc_{LM, L'M'} = \langle \de A_{LM} \de A_{L'M'}^*\rangle 
\label{covacc}
\\
&=&
\sum_{l_1 m_1 l_2 m_2}\sum_{ l_3 m_3 l_4 m_4} \cg_{l_1m_1l_2m_2}^{LM}
\cg_{l_3m_3l_4m_4}^{L'M'}
\times
\nonumber\\
&&
\biggl( C_{l_1m_1, l_3m_3}C_{l_2m_2, l_4m_4}
 + C_{l_1m_1, l_4m_4} C_{l_2m_2, l_3m_3}\biggr).
\nonumber
\ea
We assume for simplicity that posterior probability function for $\de A_{LM}$ is
normally distributed, 
\ba
 P(\vsg | A_{LM}) &\propto &
\frac{{\rm exp}\(-\frac 12 {\cal Z}^2\)
}{\det{}^{1/2} \calc}
P(A_s)
\label{prob}
\\
 {\cal Z}^2
&=& \sum_{L M L' M'} \de A_{LM}\(\calc^{-1}\)_{LM, L'M'} \de A_{L'M'}^*.
\nonumber
\ea
In order to test the above posterior probability function, we repeat some 
calculation in \cite{Picon:08} using this probability function.
We have found that the results from this probability function are in agreement with the original results.
Here, we have used a flat prior on $\Sigma$ and a Gaussian prior on $A_s$.
For the Gaussian prior, the mean and variance of $A_s$ are taken from the five-year WMAP results \cite{wmap5:para}.
Since the contributions from noncommutative geometry mainly appear in the
$\langle A_{00}\rangle$ and $\langle A_{20}\rangle$,
we restrict the multipole index $L$ of $A_{LM}$ to be less than 4.
The data that are used to compute $A_{LM}$ in eq. (\ref{delalm}) are
the five-year WMAP  foreground-reduced V2 and W1 differential assembly
temperature maps \cite{wmap5:map}.
These maps are masked using the band-limited masks in \cite{Picon:08}.
According to \cite{Picon:08}, we limit the multipole index of $c_{lm}$ and
$M_{LM}$ to be $l \leq 62$ and $L \leq 92$ respectively.
The covariance matrix $\calc_{LM, L'M'}$ in eq. (\ref{covacc}) is computed using eqs. (\ref{covat}) and (\ref{cnoi}).
Since the noncommutative geometry does not affect the cosmic evolution after the inflationary epoch,
we can use CMBEASY \cite{cmbeasy} to compute the CMB transfer function and $C_l$ by
supposing that the cosmic evolution after inflation obeys the $\Lam$CDM model whose parameters
are taken from the best fit value of the five-year WMAP results \cite{wmap5:para}.
However, recall that the amplitude of the primordial power spectrum is treated as a free parameter.
We note that the parallel computing of covariance matrices $C_{l_1 m_1, l_2 m_2}$
and $\calc_{LM, L'M'}$ can be easily implemented in the Healpix package \cite{healpix}.
We compute the posterior probability function for $\vsg$ and marginalize it over $A_s$
to obtain the marginalized posterior probability function for $\Sigma$.
The marginalized posterior probability functions for $\Sigma$ obtained from $V2$ and $W1$ maps
have a peak at negative $\Sigma$, which can occur  if $\cs^2 $ is negative.
However, we cannot use the above analysis to constrain parameters $L_s$, $\cs^2$ and $\epsilon$
simultaneously due to degeneracy among these parameters.
Since we are interested in the constraint on $L_s$,
we suppose that values of $\cs^2$ and $\epsilon$ are known.
For canonical scalar field, $\Sigma > 0$ because $\cs^2 = 1$.
Hence, we restrict ourselves to the case where $\Sigma > 0$, i.e. $\cs^2 > 0$.
In this case, the confidence intervals for
$\Sigma$ can be obtained from the areas under the curves of The marginalized posterior probability functions
in figure 1.

In order to estimate the upper bound  for $L_s$ from
the upper bound for $\Sigma$ in table 1, we further assume that
the values of the product $\epsilon\cs$ are known precisely, so that
the upper bound for $L_s$ can be written as $L_s < 1.4\times 10^4[\epsilon^2\cs^2\Mp^4]^{-1/4} {\rm Gev}^{-1}
\sim 1.4\times 10^{-28}[\epsilon\cs]^{-1/2}{\rm cm}$
at 99.7$\%$ confidence level.
It can be seen that if inflaton evolves more slowly,
the upper bound for $L_s$ will increase.
As shown in \cite{Garriga:99}, the ratio of the tensor to scalar perturbations amplitudes $r$
depends on $\epsilon\cs$.
We roughly estimate the values of $\epsilon\cs$ using the values of $r$
from the five-year WMAP results, and obtain
$L_s < 10^{-27}{\rm cm}$ at 99.7$\%$ confidence level.

\begin{figure}
\resizebox{0.45\textwidth}{!}{%
\includegraphics{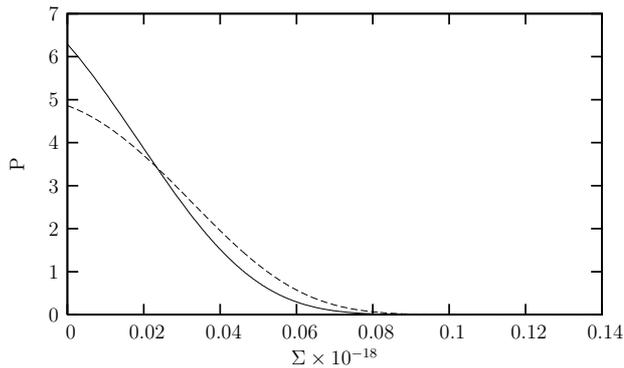}
}
\caption{
The marginalized posterior probability functions for $\Sigma$.
The solid line represents the probability function from $W1$ map, while
the long dashed line represents the probability function from $V2$ map.
Here, $\Sigma$ is restricted to be positive.
}
\label{fig::1}
\end{figure}
\begin{table}
\caption{
The upper bound for the parameter $\Sigma\times 10^{-18}$ from $W1$ and $V2$ maps.
}
\label{table1}
\begin{center}
\begin{tabular}{l|ccc}
\hline\noalign{\smallskip}

                      &
68$\%$CL &
95$\%$CL &
99.9$\%$CL \\

\noalign{\smallskip}\hline\noalign{\smallskip}

W1 & 0.024 & 0.048 & 0.070 \\
V2 & 0.029 & 0.055 & 0.078 \\
\noalign{\smallskip}\hline
\end{tabular}
\end{center}
\vspace*{5cm}  
\end{table}

\section{Conclusions}

In this work, we study the effect of noncommutative geometry on the rotational invariance of the primordial power spectrum,
and constrain the contributions from this effect using CMB data.
In the slow-roll approximation, the deviation from rotational invariance of the primordial power spectrum
due to the noncommutative effect depends on the factor
$3H^4L_s^4/16$, where $H$ is evaluated at the time when perturbation mode $k$ crosses the horizon during inflation.
Although this result is obtained using k-inflaton, it is similar to the one that is computed
from canonical inflaton field \cite{Lizzi:02}.
This implies that, in the slow-roll approximation, the effect of noncommutative geometry on primordial power spectrum
is the same for both standard inflation and k-inflation model.

Since the primordial power spectrum is direction-dependent in  our consideration,
the covariance matrix for the harmonic coefficients of the CMB temperature anisotropies has off diagonal elements.
In our case, these off diagonal elements arise from the noncommutative geometry contributions.
As is well known, this implies that statistics of the CMB anisotropies become anisotropic.
The noncommutative geometry  also  contributes to the diagonal elements of the covariant matrix
suggestting that the noncommutative contribution also modifies the amplitude of the CMB angular power spectrum.

Both contributions from noncommutative geometry  are simultaneously constrained
using five-year WMAP foreground-reduced V2 and W1 maps.
The upper bound for the quantity
$L_s[\epsilon\cs]^{1/2}$
is approximately $1.4\times 10^{-28}{\rm cm}$ at 99.7$\%$ confidence level.
If we suppose that the values of $\epsilon\cs$ are known precisely,
the upper bound for the noncommutative length scale can be written as $L_s < 1.4\times 10^{-28}
[\epsilon\cs]^{-1/2}
{\rm cm}$
at 99.7$\%$ confidence level.
This shows that if inflaton evolves more slowly,
the upper bound for $L_s$ will increase.
Estimating the values of $\epsilon\cs$ using the tensor to scalar perturbations amplitudes from the five-year WMAP results,
we obtain $L_s < 10^{-27}{\rm cm}$ at 99.7$\%$ confidence level. 

\section*{Acknowledgments}

The author would like to thank T. Souradeep for instructive conversations,
and A. Ungkitchanukit for comments on the manuscript.
He would also like to thank P. Wongjun for his checking of some parts of the calculations,
P. Burikham and A. Chatrabhuti for their comments and suggestions.
The numerical calculation was performed using the computer cluster
of Department of Physics, Kasetsart University
He acknowledges the use of the Legacy Archive for Microwave Background Data Analysis (LAMBDA)
and HEALPix package.
This work is supported by Thailand Research Fund (TRF).

\end{document}